# Sub-gap optical response across the structural phase transition in van der Waals layered α-RuCl₃

Stephan Reschke,[1] Franz Mayr,[1] Sebastian Widmann,[1] Hans-Albrecht Krug von Nidda,[1] Vladimir Tsurkan,[1,2] Mikhail V. Eremin,[3] Seung-Hwan Do,[4] Kwang-Yong Choi,[4] Zhe Wang,[5] and Alois Loidl[1,6]

[1]Experimental Physics V, Center for Electronic Correlations and Magnetism, University of Augsburg, 86135 Augsburg, Germany
[2]Institute of Applied Physics, MD 2028 Chisinau, Republic of Moldova
[3]E. K. Zavoisky Physical-Technical Institute, 420029 Kazan, Russia
[4]Department of Physics, Chung-Ang University, Seoul 06974, Republic of Korea
[5]Institute of Radiation Physics, Helmholtz-Zentrum Dresden-Rossendorf, 01328 Dresden, Germany
[6] Author to whom any correspondence should be addressed

**E-mail**: alois.loidl@physik.uni-augsburg.de

**Abstract**
We report magnetic, thermodynamic, thermal expansion, and on detailed optical experiments on the layered compound α-RuCl₃ focusing on the THz and sub-gap optical response across the structural phase transition from the monoclinic high-temperature to the rhombohedral low-temperature structure, where the stacking sequence of the molecular layers is changed. This type of phase transition is characteristic for a variety of tri-halides crystallizing in a layered honeycomb-type structure and so far is unique, as the low-temperature phase exhibits the higher symmetry. One motivation is to unravel the microscopic nature of spin-orbital excitations via a study of temperature and symmetry-induced changes. We document a number of highly unusual findings: A characteristic two-step hysteresis of the structural phase transition, accompanied by a dramatic change of the reflectivity. An electronic excitation, which appears in a narrow temperature range just across the structural phase transition, and a complex dielectric loss spectrum in the THz regime, which could indicate remnants of Kitaev physics. Despite significant symmetry changes across the monoclinic to rhombohedral phase transition, phonon eigenfrequencies and the majority of spin-orbital excitations are not strongly influenced. Obviously, the symmetry of the single molecular layers determine the eigenfrequencies of most of these excitations. Finally, from this combined terahertz, far- and mid-infrared study we try to shed some light on the so far unsolved low energy (< 1eV) electronic structure of the ruthenium $4d^5$ electrons in α-RuCl₃.

## 1. Introduction

After early reports on synthesis and structure of α-RuCl₃ [1], Fletcher *et al*. [2,3] reported on synthesis as well as on structural, magnetic, and optical characterization. The structural units of α-RuCl₃ are honeycomb layers of ruthenium, separated by two hexagonal layers of chlorine. Ru³⁺ ($4d^5$) is coordinated by Cl⁻ ions in octahedral symmetry with slight monoclinic distortion. The ruthenium layers, sandwiched between two layers of chlorine, represent strongly bonded molecular stacks, only weakly connected by van der Waals (vdW) forces. Due to the weak vdW binding energy between the molecular stacks, the structure is prone to stacking faults. Early structural studies reported on a highly symmetric $P3_112$ space group [1,2,3]. However, now it is well established that the room temperature symmetry of α-RuCl$_3$ is monoclinic with space group $C2/m$ [4,5,6], isostructural to the infinite-layer compound CrCl$_3$ at 300 K [7]. As observed in a number of layered tri-halides [7,8,9], also α-RuCl$_3$ undergoes a transition into a low-temperature rhombohedral structure with $R\bar{3}$ symmetry. This transition is located around 150 K [10,11,12,13,14,15] and is characterized by an extremely wide hysteresis extending over a temperature range of more than 100 K. A similar phase transition, induced by changes of the molecular stacking sequence, also appears in chromium tri-chloride at 240 K [7,9], in chromium tri-iodide close to 220 K [8], in chromium tri-bromide at approximately 420 K [7], and in ScCl$_3$ at ~ 950 K [16]. In the low-temperature structure, the halide ions are almost hexagonal close-packed, with an AB-type stacking sequence, in contrast to the nearly face-centered cubic ABC stacking of the high-temperature monoclinic form [7]. This monoclinic-to-rhombohedral transition can occur by a simple translational shift of neighboring layers along one direction [7]. The broad hysteresis involving a shift of neighboring two-dimensional building blocks by itself is a highly interesting phenomenon: It has been shown very recently that the twist angle between neighboring graphene layers plays a crucial role in the electronic properties [17] and can even induce superconductivity at certain twist angles [18].

Quite recently, some of these tri-halides, with their two-dimensional honeycomb lattices with weak interlayer coupling, came into the focus of modern solid-state research: α-RuCl$_3$ is a prime candidate for the realization of Kitaev physics and fractionalized quasiparticles [19], composed of itinerant and localized Majorana fermions [20,21,22]. Arguments were put forth that these exotic fractionalized excitations could exist at temperatures just above the onset of magnetic order. Indeed, recent Raman [23,24,12], neutron scattering [25,26,27,28], and time-domain THz experiments [14] provided possible experimental evidence of fractionalized excitations. However, an alternative scenario to explain the excitation continuum in this material in terms of strong magnetic anharmonicity and a concomitant magnon breakdown was proposed by Winter et al. [29]. In addition, diverse magnetic behaviors of this class of materials are of current interest: α-MoCl$_3$ exhibits strong antiferromagnetic (AFM) exchange within the planes and undergoes a dimerization transition at elevated temperatures [30]. CrI$_3$ is a ferromagnet with a critical temperature close to 60 K [31] and reveals layer-dependent ferromagnetism down to the monolayer limit [32]. This ferromagnetic (FM) behavior in atomically thin layers, termed magnetism in flatland [33], could be of importance in future data storage devices. In this compound electric control of magnetism has been documented and a reversible electric-field induced AFM to FM switching has been demonstrated [34]. All tri-halides are vdW layered materials showing unconventional phase transitions similar to experimental findings in transition-metal dichalcogenides [35].

In this manuscript, we present a detailed characterization of the exotic and strongly hysteretic structural phase transition in α-RuCl$_3$, where the stacking sequence of the vdW-coupled molecular layers is changed and the symmetry is lowered on increasing temperatures. After a magnetic and thermodynamic characterization, we provide detailed combined terahertz (THz), far-infrared (FIR), and mid-infrared (MIR) reflectivity and transmission experiments from 1 meV to 1 eV focusing on this structural phase transition. We provide experiments on reflectivity and transmittance across the phase transition to study the temperature evolution of specific phonon modes and orbital excitations.

We report on a highly complex THz spectrum, which possibly is of mixed electric and magnetic dipolar character and could indicate an underlying Kitaev physics. In addition, we document the appearance of a new electronic excitation in a very narrow temperature range just across the structural phase transition. We propose that this finding indicates changes in the electronic density of state when shifting neighboring layers of the vdW hetrostructure, as has recently been reported for graphene [17,18]. Furthermore, we contribute to the ongoing controversy about the low-energy orbital excitations in α-RuCl$_3$ and make a proposal for the spin-orbital coupled level scheme.

## 2. Methods

High-quality α-RuCl$_3$ single crystals, used in this work, were grown by vacuum sublimation. Samples were synthesized by the two groups in Seoul and in Augsburg separately. Samples grown in Seoul are characterized in detail in Refs. [11] and [27]. The characterization of the Augsburg grown samples is described below. All batches of samples behave very similar, concerning the details of the structural phase transition as well as concerning the onset of magnetic order, which is established close to 7 K. Magnetic measurements were performed with a SQUID magnetometer (Quantum Design MPMS XL) in the temperature range 1.8 K < $T$ < 400 K and in external magnetic fields up to 5 T. Standard heat capacity was investigated in a Quantum Design PPMS for temperatures 1.8 < $T$ < 300 K and in magnetic fields up to 9 T. To identify the nature of the structural and magnetic phase transitions, canonical specific-heat experiments were supplemented by a large-pulse method, applying heat pulses leading up to a 20% increase of the sample temperature. Thereby the temperature response of the sample was separately analyzed on heating and cooling. In a second-order phase transition, both responses should be equal, while significant differences are expected at first-order phase transitions. The thermal expansion experiments were again conducted with a Quantum Design PPMS equipped with a compact and miniaturized high-resolution capacitance dilatometer. For the thermal-expansion experiments, a small platelet with a thickness of 0.024 mm along the crystallographic $c$ direction was used. Despite an increased experimental uncertainty, rather thin platelets were measured to gain direct evidence of the temperature evolution of changes in the stacking sequence, in order to avoid the influence of macroscopic domains of thick bulk crystals.

The samples for the optical experiments had a typical $ab$ surface of 5 x 3 mm$^2$ and thicknesses of approximately 25 μm up to 1 mm. Thicker samples had to be used to measure transmission in the THz and reflectivity in the FIR regime. Thinner samples of various thickness were used in transmission experiments at FIR and MIR frequencies. Time-domain THz transmission experiments were performed with the wave vector of the incident light perpendicular to the crystallographic $ab$ plane using a TPS Spectra 3000 spectrometer. Time-domain signals were obtained for reference (empty aperture) and samples, from which the power spectra were evaluated via Fourier transformation. FIR and MIR reflectivity as well as transmission experiments for energies from 10 meV to 1 eV were performed using the Bruker FT-spectrometers IFS113v and IFS66v/S, with appropriate sets of sources, beam splitters and detectors. CryoVac He-flow cryostats allowed varying the sample temperature from 5 to 295 K. All IR experiments were performed with the incident wave vector perpendicular to the molecular stacks, i.e., parallel to the crystallographic $c$ direction. For all reflectivity experiments, gold mirrors were utilized as reference. To convert the reflectivity spectra into the complex dielectric permittivity $\varepsilon(\omega) = \varepsilon'(\omega) - i\varepsilon''(\omega)$ we used the Kramers-Kronig constrained variational method, which was developed by Kuzmenko [36] and is included in the RefFIT program [37]. The use of this approach allows to obtain real and imaginary part of the complex permittivity $\varepsilon(\omega)$ without the need of specific extrapolations at the low- and high-frequency edges of the measured reflectivity spectra.

## 3. Results and discussion
### 3.1 Magnetic susceptibility

Figure 1 shows the temperature dependence of the inverse magnetic susceptibility for external magnetic fields perpendicular ($H \perp c$) and parallel ($H \parallel c$) to the crystallographic $c$ direction. Each ruthenium ion is surrounded by six chlorine ions in octahedral symmetry. Due to the strong crystal-field splitting, $Ru^{3+}$ exhibits a low-spin configuration $(d\varepsilon)^5$ with five electrons in the lower $t_{2g}$ triplet with spin $S = ½$. Figure 1 documents the strong anisotropy of the susceptibility, with effective FM exchange when the external magnetic field is applied within the $ab$ planes, but with effective AFM exchange with the field perpendicular to the planes.

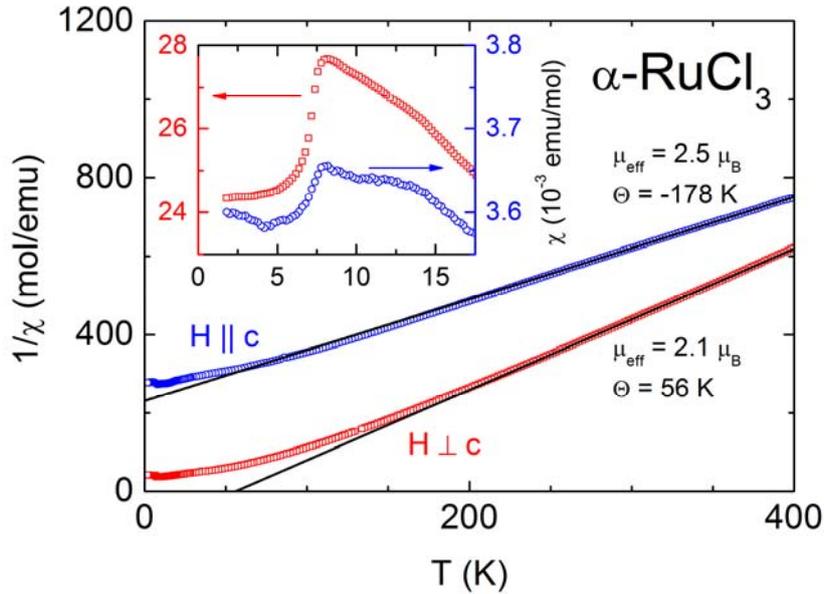

Figure 1. Inverse magnetic susceptibility of α-RuCl$_3$ as measured in magnetic fields of 1 T for temperatures between 1.8 and 400 K. Measurements are shown on heating for the external field within ($H \perp c$) and perpendicular ($H \parallel c$) to the molecular stacks. Curie-Weiss temperatures and effective paramagnetic moments are indicated for both field directions. The solid lines indicate the results of Curie-Weiss fits in a temperature range 200 - 400 K, extrapolated also to low temperatures. The inset shows the magnetic susceptibility for temperatures close to the onset of magnetic order.

In order to compare the observed magnetic susceptibility in α-RuCl$_3$ as documented in figure 1 with published results, we provide an analysis using a high-temperature Curie-Weiss (CW) behavior in the temperature regime from 200 to 400 K. This analysis yields an effective FM exchange with a CW temperature $\theta \sim 56$ K for $H \perp c$ and an effective AFM exchange of order $\theta \sim -178$ K for $H \parallel c$. The paramagnetic moment is strongly enhanced when compared to that of a spin-only $S = ½$ system with $\mu_{eff} = 1.73$ $\mu_B$. From the anisotropic effective paramagnetic moments we can calculate the g values, resulting in $g_\parallel = 2.9$ and $g_\perp = 2.4$. These values are in reasonable agreement with published results [38,39], where g values > 2 were reported for both field directions. However, susceptibility results in the work of Kubota et al. [10] were analyzed in terms of $g_\parallel = 0.4$ and $g_\perp = 2.5$. These authors concluded that the strongly anisotropic g values result from the trigonal crystal field and that the energy of the trigonal distortion is of order of the spin-orbit coupling $\lambda$.

It is evident that due to crystal-electric field (CEF) effects and strong spin-orbit coupling (SOC) the temperature dependence of the magnetic susceptibility will not follow a CW law, but has to be described by a theory modeling the paramagnetism of d electrons of complex salts taking details of CEF and SOC into consideration [40,41,42,43]. According to these models, the positive spin-orbit coupling ($\lambda > 0$) splits the $(d\varepsilon)^5$ states into a lower doublet with total spin J = ½ and an excited quartet with J = 3/2. The temperature dependence of the effective magnetic moments of this electronic configuration can be calculated as function of reduced temperature $k_B T/\lambda$, where $\lambda$ is the SOC constant. At intermediate temperatures (intermediate when compared to the strength of SOC), values of $\mu_{eff} \sim 2.5\ \mu_B$ were found [40,41], in reasonable agreement with the effective paramagnetic moments obtained from the high-temperature CW fits for $H \parallel c$. The value of the SOC constant has been determined from Raman [44] and inelastic neutron-scattering experiments [25] by measuring excitation energies of the spin-orbit split $t_{2g}$ ground state and was found to be of order 100 meV. This energy scale is small compared to the crystal field splitting ($\sim 2$ eV), but large with respect to the thermal energy corresponding to ambient temperature. Below approximately 150 K strong deviations from an extrapolated CW behavior become apparent and the CW temperatures indicated in figure 1 have to be taken as mere parametrization of experimental results.

The onset of antiferromagnetic order appears close to 7 K (see inset in figure 1), pointing towards a crystal with a well-defined low-temperature rhombohedral phase. In α-RuCl$_3$, there exists a close correlation between structural and magnetic ordering temperatures and the stacking sequence is closely linked to the onset of magnetic order. Crystals, which undergo a complete monoclinic to rhombohedral phase transition, exhibit a well-defined magnetic transition around 7 K, while crystals with a (frozen-in or super-cooled) monoclinic phase undergo a broad magnetic transition close to 14 K [45]. In a number of reports, both anomalies appear simultaneously and indicate the coexistence of both types of layering [10]. Our results show only minor, knee-like anomalies close to 14 K, documenting the absence of major stacking faults. Concerning the spin structure, the interpretation of neutron-diffraction experiments converges to a zigzag spin structure, with the spins slightly tilted from the *ab* plane and with AFM stacking along the crystallographic *c* direction [11,46].

**3.2 Heat capacity**

Figure 2 shows the heat capacity of α-RuCl$_3$ for temperatures from *T* = 2 to 300 K in zero magnetic field, as well as for an external field of 9 T, which was applied parallel to the *ab* planes ($H \perp c$). These heat-capacity experiments were performed on heating. In zero external fields, we observe a sharp lambda-like anomaly close to 7 K, indicating the transition into long-range AFM order. Small anomalies close to 11 and 14 K indicate the presence a small number of stacking faults, which can be induced by minor external pressure or even by mere manipulation of the samples [45]. In an external field of 9 T parallel to the molecular stacks the magnetic transition is fully suppressed, leaving a broad hump as a reminder of some residual entropy. This hump probably signals magnetic fluctuations, but also could correspond to the expected continuum created by fractionalized excitations [27].

It is well documented in literature that in external fields within the *ab* plane, AFM order becomes suppressed close to 7 T, resulting in a quantum critical point [47,48,49,50]. Here we focus on the structural phase transition, which appears as a small hump-like anomaly in a narrow temperature range close to 170 K. Up to magnetic fields of 9 T, the anomaly is completely independent on the external magnetic fields, which is not astonishing comparing thermal and magnetic energies. The inset of figure 2 shows the hysteretic behavior of the heat capacity in more detail on an enlarged scale, by comparing large pulse results of the heat capacity on heating and cooling. While on heating the structural phase transition is characterized by some entropy release and the heat-capacity anomaly exhibits a peak at 168 K, on cooling no anomaly can be detected at all. This is rather astonishing: In optical THz transmission [14] and in FIR as well as in MIR reflectivity experiments [13], a broad and significant two-step hysteresis on heating and cooling has been reported extending from 60 to 170 K. In the thermodynamic experiments, we only observe an anomaly on heating and this anomaly is in a very limited temperature regime between 155 and 175 K. That we miss any heat-capacity anomaly on cooling results from the measurement protocol and the extreme width of the hysteresis.

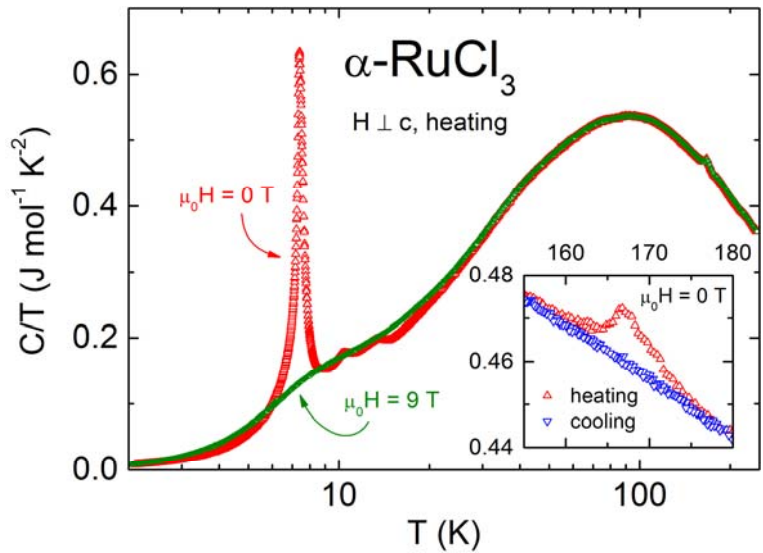

Figure 2. Temperature dependence of the heat capacity of α-RuCl$_3$. The heat capacity is plotted as $C/T$ vs. temperature $T$ on a semi-logarithmic plot. Data are shown for zero field and for an external magnetic field of 9 T with the field parallel to the molecular stacks ($H \perp c$) for temperatures between 2 and 300 K. The data in the main frame were taken on heating. The inset shows an enlarged region around the structural phase transition for heating and cooling runs using the large-pulse method.

### 3.3 Thermal expansion

To gain further insight into the nature of the structural phase transition of α-RuCl$_3$ and to understand its extreme hysteretic behavior, we performed thermal-expansion experiments, measuring length changes along the crystallographic $c$ direction on very thin samples. These experiments were stimulated by x-ray results of Park et al. [11] reporting on a two-step ordering process of this layered structure. Figure 3 documents a heating and cooling cycle of thermal expansion experiments along the $c$ direction from low-temperatures up to 220 K. Indeed, our experiments also reveal the same two-step ordering process. On cooling, the thermal expansion perpendicular to the molecular stacks exhibits conventional anharmonic behavior down to a temperature of 121 K. At this temperature, the sample abruptly shrinks along the $c$ direction and then continuously further contracts down to 55 K. On heating, the sample remains in its low-temperature phase until the $c$ axis abruptly expands at 170 K. This thermal expansion happens in a narrow temperature range and this rather abrupt structural change is monitored in the heat capacity experiments (see figure 2). The characteristic temperatures determined in this experiment, namely $T_{S1}$ = 55 K, $T_{S2}$ = 170 K and $T^*$ = 121 K are close to those reported in Ref. [11]. In the temperature regime investigated, the overall thermal expansion is of the order of 1 %, in perfect agreement with the length changes of the crystallographic *c* axis, which also amounts approximately 1 % in the respective temperature regime [11]. We would like to recall that the main hysteresis of the structural phase transition in α-RuCl$_3$ is located between $T^*$ and $T_{S2}$. However, on cooling only 50% of the changes of the *c* axis happen abruptly at $T^*$, while the remaining 50% length changes continuously evolve on further cooling down to $T_{S1}$. On heating, both lattice constants abruptly change in a narrow temperature range around $T_{S2}$.

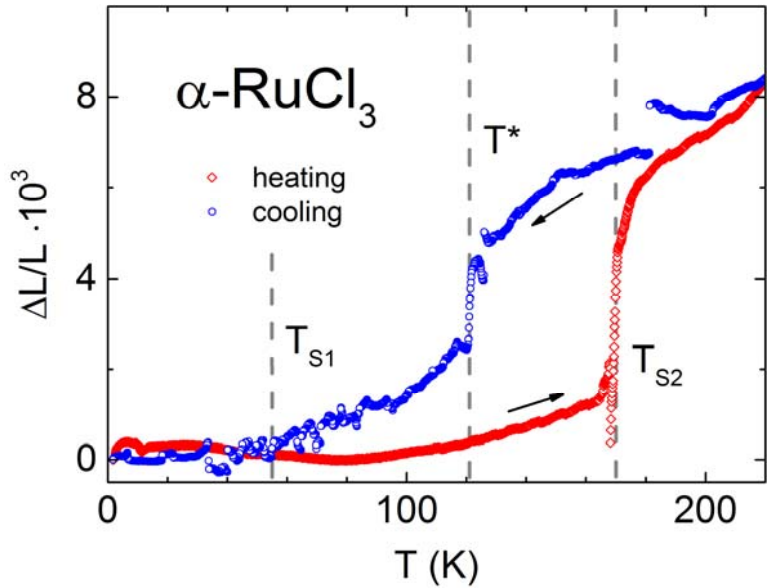

Figure 3. Temperature dependence of relative length changes $\Delta L/L$ in α-RuCl$_3$. The thermal expansion was measured perpendicular to the molecular stacks, i.e. along the crystallographic *c* direction on heating (rhombs) as well as on cooling (circles). The length changes were scaled at 0 K. Vertical dashed lines indicate the three characteristic temperatures of the two-step hysteresis behavior, where we follow the notation of Ref. [11]. The large experimental uncertainties result from the fact that we used a sample with a thickness of 24 μm only.

It is a characteristic feature of this phase transition that it resembles a two-step process on cooling when adapting the low-temperature rhombohedral phase, while it reveals a one-step process on heating, restoring the high-temperature monoclinic phase. It seems natural to assume that the temperature induced changes of the crystallographic *c* direction mainly depend on the stacking sequence and, hence, on the vdW forces between the sheets. On cooling, the changes in the stacking sequence happen partly at $T^*$ and then the stacking continuously adapts a hexagonal AB sequence at $T_{s1}$, while on heating ABC type stacking is abruptly restored at $T_{S2}$. Very recently, similar thermal expansion experiments have been performed by He et al. [15], mainly focusing on sample dependencies and hydrostatic pressure dependence. These results also show an extended hysteresis of the structural phase transition between 66 and 168 K, with a similar size of the over-all effect of the thermal expansion along the *c* axis. However, the authors of Ref. [15] did not observe the characteristic two-step hysteresis documented in figure 3.

**3.4 The dielectric response of α-RuCl$_3$ from the THz to the MIR regime**

In this work, we provide a detailed investigation of the phonon properties and of spin-orbital excitations in a broad frequency range as function of temperature: We monitor the eigenfrequencies when passing the structural phase transition, to detect significant shifts of the modes, possible splittings or the appearance of new modes indicating symmetry-lowering transitions. To provide an overview over phonon and electronic excitations in α-RuCl$_3$, figure 4 shows the energy dependence of the real [figure 4(a)] and imaginary part [figure 4(b)] of the dielectric constant from 1 meV up to 1 eV.

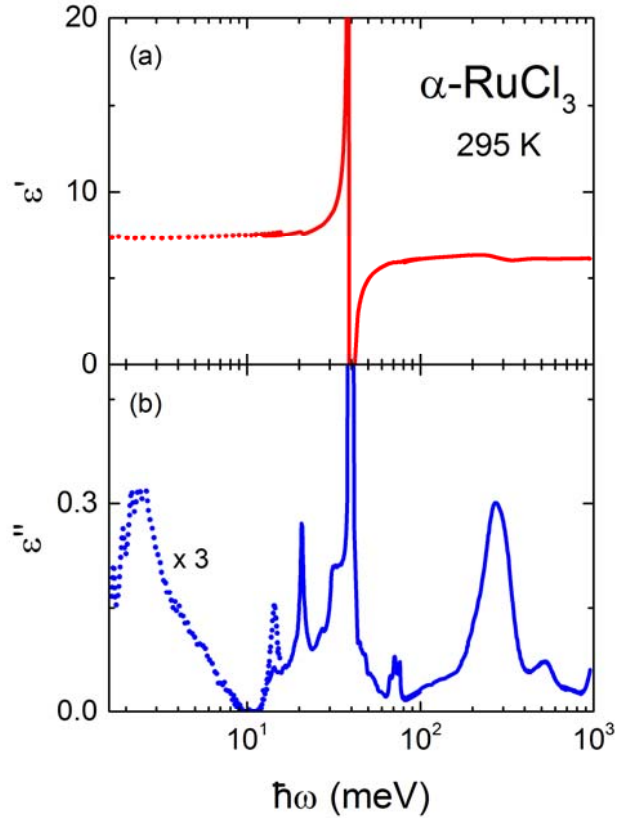

Figure 4. Low-energy broadband dielectric response of α-RuCl$_3$. Energy dependence of (a) dielectric constant $\varepsilon'$ and (b) dielectric loss $\varepsilon''$ at room temperature are plotted from 1 meV to 1 eV in a semi-logarithmic representation. All spectra were measured with the incident light perpendicular to the molecular stacks. The spectra plotted with dashed lines were derived from time-domain THz spectroscopy in transmission (< 15 meV). The dielectric loss data in the THz regime have been multiplied by a factor of three for better visibility. The solid lines result from both, Kramers-Kronig consistent reflectivity and transmission experiments in the FIR and MIR range (> 10 meV). All excitations with the exception of the phonon mode close to 40 meV are very weak and are characterized by a low dipolar weight. This becomes immediately clear by inspection of the frequency dependence of the real part of the dielectric constant $\varepsilon'$, which predominantly is determined by electronic contribution beyond 1 eV. The increase in the dielectric constant $\Delta\varepsilon'$ derived from the strongest phonon modes amounts ~ 1. The strongest spin-orbital excitation close to 300 meV, in the real part is visible as tiny anomaly and marginal step, only. In Refs. [13,51] this anomaly in the complex dielectric constant has been assigned as charge gap.

The real part of the dielectric response below 1 eV shows only one phonon-like excitation close to 40 meV. The magnitude of the dielectric constant shown in figure 4(a) is dominated by electronic contributions beyond 1 eV. Only the phonon mode close to 40 meV carries considerable dielectric strength $\Delta\varepsilon \sim 1$ [13], while all the other modes have negligible dipolar weight. In the imaginary part of the dielectric response, a number of excitations from 1 meV up to 1 eV are visible, which will be discussed below. A similar spectrum of the dielectric response has been published previously by Reschke et al. [13] and represents the outcome of an analysis of combined THz, FIR and MIR results, analyzing both, transmission and reflectivity measurements using a series of crystals of different thickness ranging from ~ 25 μm to ~ 1 mm. The absorption coefficient at FIR and MIR frequencies has also been published by Hasegawa et al. [51], including first-principle calculations of the phonon properties of α-RuCl$_3$. Biesner et al. conducted a detailed study of the pressure dependence of the conductivity spectrum up to 1.5 eV [52].

Roughly speaking, concerning the dielectric response of α-RuCl$_3$, if at all, remnants of Kitaev physics are expected below 10 meV, phonon modes and multi-phonon absorptions will appear between 10 and 100 meV, while beyond 100 meV, on-site spin-orbital excitations between different electronic levels of the ruthenium $d^5$ electrons are expected. One also should have in mind that the question concerning the size of the electronic band gap is not finally solved. α-RuCl$_3$ is a good insulator and the majority of experimental and theoretical results point towards an optical gap of order 1 eV [53], but energy gaps of 1.9 eV determined from photoemission studies are also reported [54]. However, there are some arguments in favor of a much smaller charge gap of order 200 - 300 meV [13,51]. Figure 4 documents the frequency dependence of dielectric constant and dielectric loss of a good insulator, with no apparent free charge carriers at room temperature. Any significant contribution of dc conductivity would result in a continuous increase of the dielectric loss towards low energies, which cannot be detected, even not down to 1 meV and even not at the highest temperatures investigated [figure 4(b)].

In the THz regime, at room temperature we detect a rather weak and broad excitation close to 2.5 meV probably of spin-orbital nature, as well as a phonon excitation close to 15 meV. In the two-dimensional honeycomb-type tri-halides, phonon modes so far were detected between 10 and 50 meV: early IR experiments on α-RuCl$_3$ were published by Fletcher et al. [3] and Guizetti et al. [55]. Later on, combined reflectivity and transmission results were reported by Little et al. [56] and finally, a detailed analysis of phonon excitations was presented by Reschke et al. [13] and by Hasegawa et al. [51]. Electronic excitations dominate the frequency dependent complex dielectric constant beyond 100 meV. These excitations are very weak, as they are parity and spin forbidden. They can gain dipolar intensity only via $p$-$d$ coupling or via coupling to lattice vibrations. Hence, these excitations are almost invisible in reflectivity and have to be studied in transmission geometry. Sandilands et al. [44] have studied spin-orbit excitations in detail. In good agreement with earlier results [55,57], these authors identified excitations from the $t_{2g}^5$ ground state to excited spin-orbit states of $t_{2g}^4 e_g^1$ character close to 0.3, 0.5 and 0.7 eV. Reschke et al. [13] and Hasegawa et al. [51] performed closely related studies with similar results. The latter authors also discussed the possibility that the lowest excitation close to 300 meV indicates the onset of the optical band gap.

In what follows we discuss the temperature evolution of the spectra in the different frequency regimes, i.e. at THz, FIR and MIR frequencies and specifically when passing the structural phase transition. This will allow characterizing the structural phase transition, which lowers the crystal symmetry on increasing temperature. It also could help to learn more about the nature of the different spin-orbital and phonon-like excitations.

**a) Excitations in the THz regime: Remnants of Kitaev physics?**

Figure 5 shows the frequency dependence of the dielectric loss of α-RuCl$_3$ for temperatures between 10 and 200 K. The well-defined mode close to 2.5 meV, which is rather temperature independent and the appearance of a second excitation band around 7 meV in the low temperature phase are striking and there seems to be no straight-forward explanation. These excitations carry very low optical weight and can only be observed in transmission experiments on relatively thick (~ 1mm) samples. With the focus on the possible occurrence of a continuum due to fractionalized excitations, Wang et al. [14] discussed this frequency regime in detail. To avoid possible interference with spectral changes driven by the structural phase transition, the focus of the analysis of the THz spectra in Ref. [14] was on temperatures below 60 K. Detailed investigations of the THz regime between 1 and 6 meV have also been published by Little et al. [56] and by Shi et al. [58].

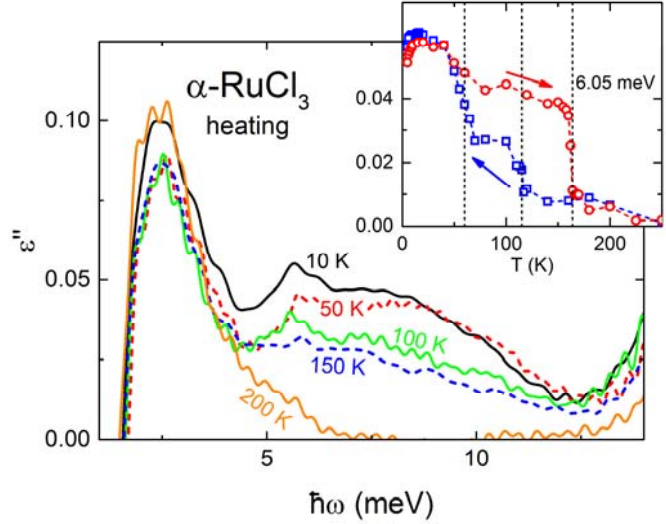

Figure 5. Energy dependence of the dielectric loss in α-RuCl$_3$ in the THz regime. These data have been taken on heating for a series of temperatures between 10 and 200 K. The inset displays the temperature dependence of the dielectric loss at 6.05 meV for a heating and cooling cycle and visualizes the two-step nature of the structural phase transition on cooling. In this energy regime, the THz loss is high in the rhombohedral low-temperature phase and approaches zero in the monoclinic high-temperature phase.

Here we specifically discuss the THz excitations to find experimental signatures of fractionalized excitations of the Kitaev-type spin-liquid phase. To elucidate the microscopic origin, we specifically focus on the observed excitations when passing the structural phase transition. At 10 K just above the onset of long-range magnetic order, the spectrum consists of a well-defined peak at 2.5 meV with a width ~ 2 meV, which shows only minor temperature induced changes up to room temperature, even when passing the structural phase transition. The steep and well-defined onset of dielectric loss at low energies is compatible with the existence of a gap of ~ 1 meV in the excitation spectrum. In addition, we observe a much broader excitation, which extends from 4 meV up to 12 meV and exists in the low-temperature phase only. The increase of the loss towards high frequencies beyond 10 meV stems from the phonon mode, which is located close to 15 meV [figure 4(b)] and has also been observed by neutron scattering (see Supplemental Information of Ref. [27]). At first sight, the low-frequency mode remains unassigned. With the SOC constant of λ ~ 100 meV [25,44] in mind, it seems unlikely to expect spin-orbital excitations below 10 meV, even assuming that the spin-orbital derived quartet and doublet states are further split into three Kramers doublets by a trigonal distortion of the octahedral CEF [59]. Low-frequency spin-orbital excitations have been observed in a number of spin-orbital coupled 3$d$ transition-metal chalcogenides [60,61,62,63]. However, even a trigonal distortion of the octahedral crystal field would result in a pure doublet. Lower symmetry environment or external magnetic fields are needed to further split this doublet, which could explain this low-lying excitation. It seems very unlikely that any canonical scenario of a coupled spin-orbit level scheme will explain the experimentally observed spectra of α-RuCl$_3$. An alternative scenario is offered below.

As documented in figure 5, significant temperature-induced spectral changes appear at excitation energies from 5 to 12 meV, where a broad hump evolves on decreasing temperatures. This additional excitation only exists in the low-temperature rhombohedral phase. On heating, this excitation reveals only minor changes up to 170 K, where it abruptly disappears and is absent in the high-temperature monoclinic phase. Phonon intensity close to 7 meV has been identified in neutron-scattering experiments (see Supplemental Information of Ref. [27]). However, in these neutron experiments the excitation intensity is increasing with increasing temperature and strongest at 200 K, contrary to our results. If we follow the temperature evolution of the dielectric loss close to 6 meV, we find the characteristic hysteresis due to the phase transition, which is documented by the thermal expansion experiments (figure 3): Only minor changes appear below 60 K, while above 170 K this excitation is almost zero within experimental uncertainty (see figure 5). For frequencies between 1 – 6 meV and temperatures from 2 K to room temperature, also Little et al. [56] reported detailed conductivity spectra. Their results are similar to the findings shown in figure 5. There is a well-defined cut-off frequency of 1 meV, an only weakly temperature dependent peak at 1.5 meV and an increasing conductivity towards 6 meV, which, however, does not become completely suppressed towards high temperatures. The THz conductivity spectra published by Shi et al. [58] show a broad hump close to 4 meV, which very weakly depends on temperature and external magnetic field. The broad conductivity hump still exists at 7 T and even at room temperature. As the observed spectral features of α-RuCl$_3$ mainly are determined by the structure of the molecular layers, these observed differences cannot be explained by differences in the stacking sequence and at present, it is unclear to what extent these THz responses reflect fractionalized excitations of Kitaev-type spin liquids or rather vibrational excitations in this complex stacked layer compound.

We also try to compare our THz results with those deduced from Raman spectroscopy [12,23] and from inelastic neutron scattering experiments [25,27,64]. Utilizing Raman scattering techniques, Sandilands et al. [23] and Glamazda et al. [12] report on a low-temperature continuum extending up to ~ 25 meV, certainly much wider in energy than the THz observations of this work documented in figure 5. This fermionic intensity exhibits a significant temperature dependence, but still is expected to be observable at room temperature [24]. In neutron scattering, at low temperatures the absence of structured scattered intensity below 2 meV confirmed a gap in the magnetic excitation spectrum [25]. In addition, a broad peak was observed, centered at approximately 5 to 8 meV [25,27,64], which is not too far from the findings reported in figure 5. This peak has maximum intensity at low temperatures and continuously decreases on increasing temperature. A correlation with the structural phase transition has not been investigated. All these Raman and neutron scattering studies cited above, in the paramagnetic phase provide no experimental evidence for a peak-like structure close to 2 meV, as shown in figure 5.

That indeed, the dielectric loss spectrum as documented in figure 5 indicates some remnants of Kitaev physics, stems from a recent theoretical work from Bolens et al. [65]. These authors calculated the low-frequency optical conductivity of Kitaev materials assuming an interplay of Hund's coupling, spin-orbit coupling, and crystal-field effects. Using realistic parameters for α-RuCl$_3$ they find a low frequency peak of magnetic dipolar response close to 1 meV, a dominant peak of electric dipolar nature close to 6 meV and a low-frequency gap value close to 0.3 meV. This predicted spectrum bears striking similarities with the loss spectrum shown in figure 5 just above the onset of magnetic order at 10 K, with a gap-like feature close to 1 meV and two peaks located at 2.5 and 7 meV. Certainly a number of open questions remain, which have to be explained theoretically: The low-frequency peak has almost no temperature dependence up to room temperature. The 7 meV peak only exists in the low-temperature phase and is absent in the high-temperature monoclinic phase.

Finally, nothing is known about a possible dynamic magneto-electric coupling in this material. However, weak and strongly anisotropic magneto-electric effects at audio frequencies were reported recently [66].

**b) Excitations in the FIR regime: Phonons and orbital excitations**

From published experimental work, so far it is evident that the phonon modes are determined mainly from the $D_{3d}$ symmetry of an isolated molecular stack. We would like to recall that assuming $D_{3d}$ symmetry in IR spectroscopy two $A_{2u}$ and three $E_u$ modes are expected [51,67]. In clear distinction, in rhombohedral $R\bar{3}$ symmetry three $A_u$ and three $E_u$ modes, while in the high-temperature monoclinic $C2/m$ phase four $A_u$ and five $B_u$ modes should be observed [68]. To search for these modes and for the possible appearance of new modes at the structural phase transition, we performed new detailed and systematic transmission experiments. Reflectivity spectra of α-RuCl$_3$ for energies between 20 and 70 meV and temperatures between 5 and 290 K were published previously in Ref. [13]. The FIR reflectivity is dominated by one reststrahlen band located around 40 meV. The maximum reflectivity of this band changes considerably with temperature, but the limiting frequencies, which correspond to the transverse and longitudinal optical branch, remain constant within experimental uncertainties. This latter fact signals that the optical weight of this mode does not change significantly, however, the reflectivity does. In addition, the temperature dependence of the maximum reflectivity of this reststrahlen band was documented in Ref. [13] for a complete hysteresis cycle. On heating, the reflectivity changes abruptly close to 170 K, in the temperature range where the rhombohedral phase completely transforms into the monoclinic high-temperature phase in a narrow temperature interval. On cooling, the well-known two-step hysteresis was observed, documenting that the temperature dependence of the reflectivity is closely linked to changes of the crystallographic $c$ axis when crossing the structural phase transition. We would like to recall, that in the heating run, the reflectivity indicates a step-like increase by almost a factor of three at 170 K [13].

In this work, we will focus on the absorption as measured in the FIR regime. Figure 6 shows the transmission as measured on heating for energies from 10 to 80 meV and for temperatures from 10 to 200 K. In the complete FIR regime, we find a series of rather strong absorption bands. IR spectroscopy results on a variety of isostructural tri-halides have been published by Emeis et al. [69], Bermudez [67], and Borghesi et al. [70]. In all these experiments, no one-phonon excitations of $E_u$ symmetry are reported below 10 and beyond 50 meV. In addition, taking into account the first-principle calculations of phonon modes of α-RuCl$_3$ by Hasegawa et al. [51] and of non-magnetic RhCl$_3$ from Wolter et al. [50], we expect no one-phonon excitations for energies above 50 meV. On the other hand, taking existing information so far, we do not expect any orbital or spin-orbital excitations below 100 meV. In the most detailed quantum chemistry calculations [71] the excitations between the spin-orbital split ground state are well beyond 100 meV in agreement with experimental results [25,44].

Figure 6 provides experimental evidence that - with the exception of small shifts and concomitant changes of the excitation width of the 21 meV phonon - no specific symmetry changes of phonon modes, when crossing the structural phase transition, appear. Hence, it seems justified to assume that $D_{3d}$ symmetry of a single molecular stack governs the lattice dynamics of α-RuCl$_3$. In this case and in the geometry chosen in the experiment documented in figure 6, we expect three $E_u$ modes. $A_{2u}$ modes can only be detected by a misalignment of the sample or by misaligned domains within the sample. Theoretically, IR active modes of $E_u$ symmetry are expected at 20.7, 33.7 and 38.3 meV, while $A_{2u}$ modes have been predicted at energies of 15.1 and 35.2 meV [51]. The arrows in figure 6 indicate the experimentally observed absorption bands located at 14.7 and 34.2 meV for the $A_{2u}$ modes and at 21.3, 32.0 and 39.2 meV for the $E_u$ modes. All these mode energies are close to the theoretical predictions. We also detect one extra mode close to 31 meV, which cannot be assigned. The small spike close to 27 meV probably corresponds to a two-phonon process. One problem is that three of these modes are close to the energy of the mode with the strongest absorption and it is unclear if their assignment is reliable (figure 6). However, it also seems possible that the real symmetry is lower than that of the assumed $D_{3d}$ symmetry and that in total 6 modes are expected in the rhombohedral and 9 modes in the monoclinic phase and that the majority of these modes are missed due to very low dipolar weight.

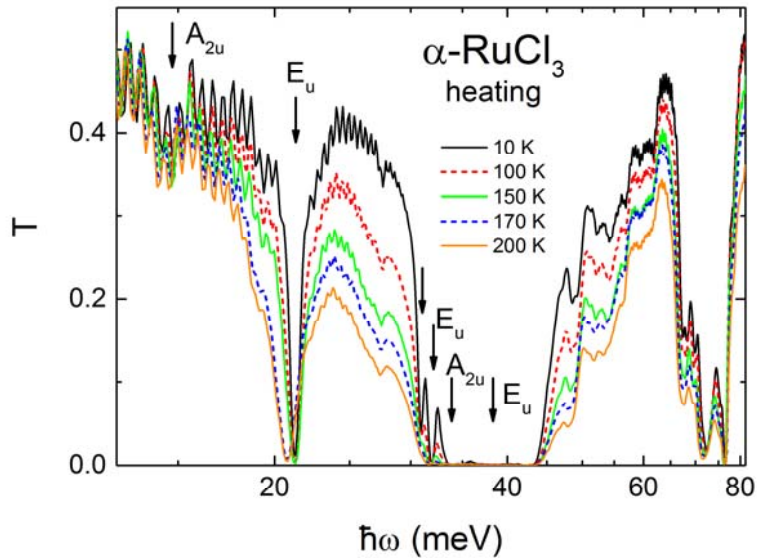

Figure 6. Energy dependence of the transmission $T$ in α-RuCl$_3$ for photon energies from 10 to 80 meV. Transmission spectra are shown on a semi-logarithmic plot for a series of temperatures between 10 and 200 K across the structural phase transition. In these experiments, the transmission has been measured on heating. Above 50 meV, multi-phonon processes dominate the transmission. The arrows indicate one-phonon excitations in $A_{2u}$ and $E_u$ symmetry in accordance with theoretical predictions of Ref. [51]. In the geometry chosen for these experiments, with the incident wave perpendicular to the $ab$ plane, we expect only three $E_u$ modes assuming $D_{3d}$ symmetry of a single molecular stack (see text). $A_{2u}$ modes can only become visible if the sample is slightly misaligned or from misaligned domains within the sample. The strong absorption band close to 70 meV is a matter of dispute and it is unclear if it can be explained in terms of multi-phonon scattering processes (see text). The mode at 31 meV remains unexplained and possibly corresponds to the lower symmetry of the real three-dimensional crystal lattice.

If one-phonon processes only exist below 50 meV, the significant absorption, which is visible in α-RuCl$_3$ beyond 50 meV (figure 6) has to be ascribed to multi-phonon processes or to orbital excitations. We think that multi-phonon processes cannot explain the dominant absorption band close to 73 meV. Two-phonon excitations, which extend to higher energies, are rather weak [67,70]. For example, two-phonon modes reported for isostructural CrCl$_3$, indeed have been observed in transmission experiments [67], but appear as very weak absorption bands only. Hence, one should take into consideration that this strong absorption band at 73 meV stems from an electronic transition. As mentioned earlier, in octahedral symmetry and strong crystal field, the ground state will be a t$_{2g}$ triplet state with spin S = ½, which under SOC further splits into a low-lying doublet ($J$ = 1/2) and an excited quartet ($J = 3/2$). Deviations from cubic crystal-field symmetry will further split these states into three Kramers doublets and we speculate that the absorption band close to 73 meV corresponds to a transition from the ground state to the lowest excited doublet. In literature there exist only very few reports of excitations of ruthenium ions with $d^5$ configuration in octahedral fields. Geschwind and Remeika [59] measured $Ru^{3+}$ in $\alpha$-$Al_2O_3$, which substitutes Al and is located in an octahedral crystal field with trigonal distortion. In these experiments, the level splitting due to deviations from cubic symmetry was of order 100 meV, not too far from the excitation frequency observed in our experiments. Estimates of the level splitting due to the trigonal distortion in $\alpha$-$RuCl_3$ range from 12 meV from an analysis of x-ray absorption spectroscopy [72], to 70 meV from quantum chemistry calculations [71] and even up to ~ 100 meV, the order of spin-orbit coupling, from an analysis of the anisotropy of the g-values [10]. Further experiments and model calculations will be necessary to identify the nature of this 73 meV excitation in $\alpha$-$RuCl_3$.

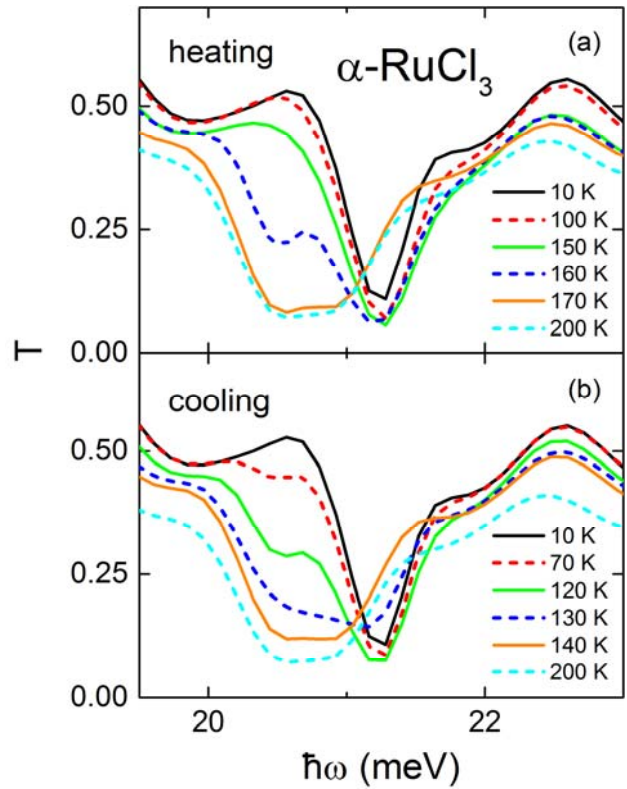

Figure 7. Transmission spectra in $\alpha$-$RuCl_3$ close to the phonon mode at 21 meV for a series of temperatures. The energy dependent transmission is reported on crossing the structural transition during heating (a) and cooling (b). Note the extended two-phase region on cooling, which extends at least from 70 to 130 K.

Figure 7 documents the sensitivity of the specific phonon mode close to 21 meV on structural changes of $\alpha$-$RuCl_3$ when passing from the low-temperature rhombohedral to the high-temperature monoclinic phase. At first sight (see the transmission documented in figure 6), the phonon modes undergo no significant changes from room temperature down to the lowest temperatures. Specifically, no new modes become visible above experimental uncertainty [13], although the very weak dipolar weight of some of these vibrations and the strong multi-phonon scattering significantly hamper the identification of new modes. Nevertheless, the phonon mode close to 21 meV reveals significant changes clearly beyond experimental uncertainties when passing the structural phase transition.

Figure 7 shows some representative scans for heating [figure 7(a)] and cooling [figure 7(b)]. The narrow mode at 21.2 meV is characteristic for the low-temperature rhombohedral phase, while a relatively broad excitation at 20.7 meV is the fingerprint of the high-temperature monoclinic phase. In a wide temperature regime, these two excitations coexist, documenting the first-order character of the structural phase transition, with a broad heterogeneous coexistence regime of both phases. On heating, the narrow mode exists up to at least 150 K and then rapidly transforms into the excitation characteristic for the monoclinic phase, which is fully developed at 170 K. On cooling, traces of the high-temperature mode still are visible down to 70 K, fully compatible with the hysteresis documented in the thermal-expansion experiments shown in figure 3. On cooling the two-phase region extends from 130 down to 70 K, while on heating the two-phase region is limited to a narrow temperature range around 160 K. It seems important to note that an isosbestic point appears close to 21 meV indicative of a well-established two-phase regime. This phonon mode turns out to be an ideal probe for the crystallographic symmetry and the stacking sequence of the samples under consideration. A narrow line signals the rhombohedral phase with AB stacking. A broad line at slightly lower frequencies is the fingerprint of the monoclinic high-temperature phase. A splitting of this mode is a characteristic feature of samples with a large amount of stacking faults and mixed monoclinic and rhombohedral phases.

**c) Excitations in the MIR regime: Orbital excitations**

Finally, we discuss transmission and reflection experiments on α-$RuCl_3$ in the MIR regime. In IR spectroscopy, these excitations naturally will be very weak because they are both, spin and parity forbidden and usually can only be observed in transmission geometry. Figure 8 shows the transmission spectra between 100 meV and 1 eV as observed at 10 K and at 295 K. At first sight, at both temperatures one can observe three characteristic transmission bands. This interpretation is in accord with earlier reports [53,13,51]. However, a closer inspection indicates a fourth absorption band just above 400 meV. This is indicated by a strong asymmetry of the transmission maximum between first and second absorption band. The transmission profiles significantly sharpen on decreasing temperatures indicating strongly decreasing damping. The transmission of the low-frequency band approaches zero, indicating that in this energy regime the sample was slightly too thick.

The transmission spectra have been fitted with Lorentzian profiles. In figure 8 the fits are indicated as solid lines. At room temperature, we observe four electronic absorption bands centered at eigenfrequencies of 277, 474, 542 and 752 meV and with a characteristic damping of 15, 115, 137 and 276 meV, respectively. At 10 K these absorption bands are located at 294, 465, 536 and 740 meV with damping factors of 11, 60, 111 and 224 meV. On cooling, the lowest frequency band undergoes a significant blue shift, a hallmark of electronic transitions coupled to vibronic states. There are only minor temperature-induced shifts for the absorption bands close to 465 and 536 meV and a conspicuous small red shift for the high-frequency band. On decreasing temperatures, the damping of the lowest two absorption bands decreases up to a factor of two, while the damping of the higher frequency bands decreases by approximately 20 % only. It is interesting to note that the intensity of these absorption bands are rather temperature independent and on lowering temperatures remain constant within less than 10 %. These temperature induced shifts and broadening effects are rather continuous and do not seem to be significantly influenced by the structural phase transition. We conclude that the symmetry of the crystal electric field of a single molecular stack mainly determines these on-site absorption bands. Figure 8 also signals a strong increase of the transmission below 200 meV. In Ref. [13] this absorption edge has been interpreted as the band edge for electronic transitions across the optical band gap and would be consistent with band gaps as derived from temperature dependent dc resistivity results [57]. This optical gap also shows an enormous blue shift on decreasing temperatures. However, this interpretation is at odds with a number of reports claiming gap values of the order 1 to 2 eV [53,54].

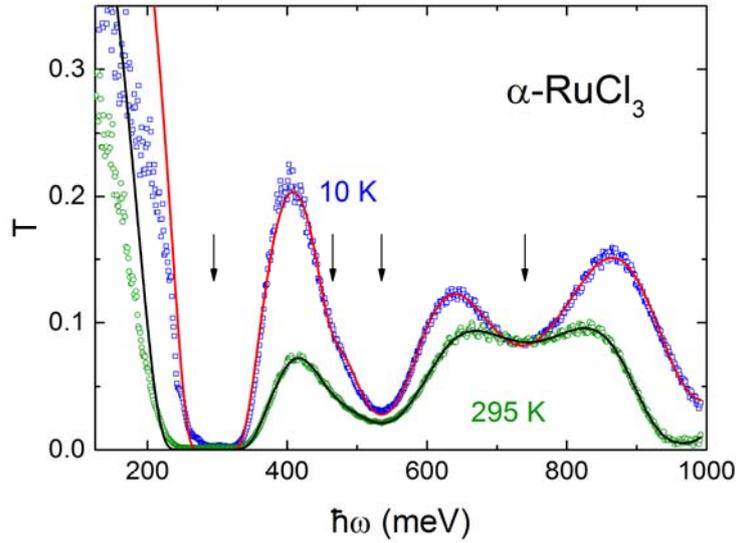

Figure 8. Energy dependence of the transmission $T$ in α-RuCl$_3$ for photon energies from 150 meV to 1 eV. As the eigenfrequencies of the transmission bands only weakly depend on temperature, transmission spectra are shown at 10 and 295 K only. The solid lines represent fits of the transmission profile using Lorentzian-type fits. The vertical arrows indicate excitation energies at 10 K for four bands that have been assumed in these fits (see text). The extremely strong decrease of the transmission close to 200 meV, in Ref. [13] was analyzed in terms of a direct allowed transition across the optical band gap.

So far, in optical spectroscopy three orbital excitations were identified located close to 300, 500 and 700 meV [44,13,51]. They were assigned to excitations of the Ru $d^5$ electrons from the $t_{2g}^5$ ground state to excited spin-orbit states of $t_{2g}^4 e_g^1$ character. The excitation between the spin-orbital split ground state has only been detected by Raman spectroscopy close to 145 meV [44] and was not identified by any IR experiment. A spin-orbit splitting of the order 195 meV was deduced from inelastic neutron-scattering experiments [25]. Neither Raman nor neutron experiment were able to identify a further splitting of this doublet to quartet transition, due to the trigonal distortion of the crystal field. From x-ray absorption spectroscopy, this splitting was estimated to be of order 12 meV [72], while the analysis of the g-values from magnetic susceptibility experiments resulted in much larger values [10]. Based on the present experiments and on our analysis as detailed above, we identify a total of 5 orbital excitations below 1 eV: At room temperature these are located at 73, 277, 474, 542 and 752 meV. However, we would like to state that this interpretation is by far not consistent with recent ab-initio calculations, where these $t_{2g}^5 \rightarrow t_{2g}^4 e_g^1$ excitations have been predicted for energies well beyond 1 eV. Yadav et al. [71] have reported detailed electronic structure calculations and corresponding energies of electronic excitations. These calculations do not support the interpretation of the excitations between 100 meV and 1 eV as being due to $t_{2g} \rightarrow e_g$ transitions, which are computed at energies significantly beyond 1 eV. According to these quantum-chemistry calculations, only two excitations should be observed at low energies. Depending on details of the model and the assumed symmetry, the spin-orbit splitting is of order 200 meV, with a trigonal splitting of 50 meV [71].

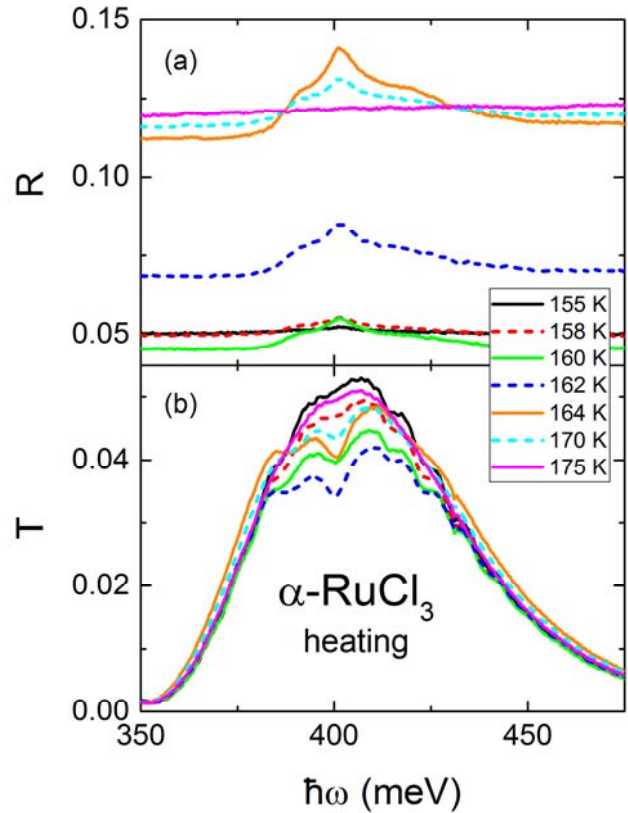

Figure 9: (a) Mid-infrared reflectivity R and (b) transmission T in α-RuCl$_3$ around 400 meV. Reflectivity and transmission are shown for a series of temperatures from 155 to 175 K on heating. In this temperature regime, the crystal abruptly transforms from the low-temperature rhombohedral into the high-temperature monoclinic structure. The transmission is identical in the high- and low-temperature ordered phases. Please note that in this narrow temperature range the MIR reflectivity changes by a factor of 2.5 between high and low-temperature phases, while the transmission is hardly affected.

Here we want to document the occurrence of a new electronic transition close to 400 meV, which appears only in a very narrow temperature range on heating, when the crystallographic structure changes from the low-temperature rhombohedral into the high-temperature monoclinic phase. This electronic excitation is clearly visible in reflectivity [figure 9(a)] as well as in transmission [figure 9(b)] experiments, but obviously has very low optical weight. This electronic excitation only exists (or is only IR active) in the heating cycle in the narrow temperature range between 155 and 175 K, exactly in the transition region where the rhombohedral phase transforms into the monoclinic one (see figure 3). It is invisible in reflectance as well as in transmission for higher and lower temperatures in both fully ordered phases. One plausible reason for its occurrence just in the phase-transition regime could be that this transition is fully spin and/or parity forbidden in the ordered phases and hence is invisible in IR experiments, but becomes disorder-allowed just at the phase transition where the stacking sequence changes. In this structural phase transition, the stacking sequence changes most likely by a continuous shift of neighboring molecular stacks. A highly intriguing explanation would be that the electron density of states changes when neighboring stacks are shifted. A similar observation, namely that the electron density of a hetero structure strongly depends on the twisting angle of neighboring layers has recently been reported for graphene [17,18]. Of course, much more experimental and theoretical work is needed to document such a behavior.

Figure 9 also provides experimental evidence for significant differences in changes of transmission and reflectance, when passing the structural phase transition. Figure 9(b) documents that the transmission is identical within experimental uncertainty in the high and low temperature phases, but reveals a significant decrease in the transition region. The excitation in reflectivity becomes apparent only when crossing the phase transition [figure 9(a)]. However, this figure impressively documents the significant change in the overall reflectivity at MIR frequencies, which was documented earlier [13]. The reflectivity in the monoclinic phase with cubic ABC stacking is by a factor of 2.5 higher than the reflectivity in the rhombohedral phase with hexagonal AB stacking sequence. As outlined above, this dramatic change in reflectivity is a characteristic feature accompanying the change in the stacking sequence of α-RuCl$_3$ and can be observed from 1 meV up to 1 eV. It possibly indicates additional broadband reflectivity or absorption processes at the surface or at the interfaces of the vdW coupled molecular layers.

## 4. Conclusions

In this work, we provide a detailed study of phononic, spin-orbital and orbital excitations in α-RuCl$_3$ by broadband optical experiments in the THz, FIR and MIR frequency range. We specifically focus on the eigenfrequencies of these excitations when crossing the structural phase transition, which is unusual in the sense that on increasing temperatures a high-symmetry rhombohedral phase is followed by a lower symmetry monoclinic phase. The optical studies are complemented by detailed measurements of magnetic susceptibility, heat capacity and thermal expansion. It seems now well established that the structural phase transition in α-RuCl$_3$ mainly signals changes in the stacking sequence of the molecular layers, but that the crystallographic structure of the molecular Cl-Ru-Cl layers remain essentially identical. By comparing with related tri-halide compounds, it seems plausible to assume that an AB stacking in the rhombohedral phase at low temperatures is followed by an ABC stacking sequence in the monoclinic high-temperature structure. The phase transition is of strongly first order and has an unusual width extending between T$_{s1}$ = 55 K and T$_{s2}$ = 170 K. In addition, the cooling cycle of the hysteresis exhibits a clear two-step behavior, where the c axis shrinks between T$_{s2}$ and a characteristic temperature $T^*$ = 112 K by approximately 50 % and then continuously further relaxes towards its low-temperature equilibrium value at a temperature $T_{S1}$. On heating, the low-temperature rhombohedral structure changes abruptly into the high-temperature monoclinic phase at a characteristic temperature $T_{S2}$.

Our thermal expansion experiments are in very good agreement with the structural details determined from x-ray diffraction [11]. These structural changes concomitantly happen with changes of the stacking sequence. The temperature dependent heat capacity of α-RuCl$_3$ shows only a small anomaly at the upper critical temperature of the hysteresis, $T_{S2}$. The entropy gain at the phase transition is marginal compared to the phonon contributions and probably indicates that entropy only plays a minor role as the driving force of this exotic phase transition. In the temperature dependence of the magnetic susceptibility, we find no traces of the structural phase transition. The magnetic susceptibility is strongly anisotropic, signals the onset of antiferromagnetic order below 7 K and is compatible with the assumption of a spin-orbit split ground state with a lower J = ½ doublet and an excited J = 3/2 quartet. In the optical experiments we document that this prominent two-step relaxation of the c axis is mirrored in the reflectance and transmittance at all frequencies investigated, namely at THz, FIR and MIR frequencies. In clear distinction, the eigenfrequencies of phonon modes as well as of spin-orbital excitations reveal no significant influence of the structural phase transition, with the exception of a broad excitation evolving in the THz regime in the rhombohedral phase and a clear shift of the phonon mode close to 21 meV.

Focusing on the phononic and electronic excitations below 1 eV reported in this work, a number of ambiguities remain. In the THz regime at low temperatures, we found clear experimental evidence for two excitations close to 2.5 and 7 meV, including a gap-like cut-off frequency of 1 meV (figure 5). These results bear some similarities with recent calculations of polarization and spin correlation functions on the Kitaev honeycomb lattice [65] and, hence, could signal some remnants of Kitaev physics. There have been published Raman [23,24], neutron scattering [25,27,64] and THz results [14,56,58] focusing on the low-energy excitation spectrum in α-RuCl$_3$. All these studies report on an existing low-energy continuum. However and to be honest, neither from the energy range nor from the temperature dependence, these reported results bear significant common universalities, which unambiguously point towards a fractionalized excitation spectrum of Majorana fermions. Of course, one possible explanation could be that details of the microscopic interactions are strongly sample dependent and e.g. depend on stacking sequence of the molecular stacks and hence, on the phase transition. Much more theoretical and experimental work will be needed to unravel the low-energy excitation spectrum in this fascinating compound.

The over-all phonon properties α-RuCl$_3$ can rather well be described by assuming $D_{3d}$ symmetry of a single molecular stack. However, we have to admit that a detailed analysis is hampered by the very low dipolar weight of most of the excitations and the strong temperature dependent multi-phonon scattering. At room temperature, we assigned two $A_{2u}$ modes at 14.7 and 34.2 meV and three $E_u$ modes 21.3, 32.0 and 39.2 meV. A mode close to 31 meV remains unassigned. Finally, we observe a strong absorption band at 73 meV, which according to Ref. [51] should be interpreted as multi-phonon absorption. We argue that for a two-phonon scattering process, the experimentally observed absorption is far too strong and we favor an interpretation in terms an electronic transition.

The interpretation of the electronic transitions in α-RuCl$_3$ below 1 eV is far from being solved. According to this work, five spin-orbital excitations were identified, which are located at 73, 277, 474, 542 and 752 meV. Two of these excitations could correspond to transitions between the SOC ground state with additional trigonal splitting. This scenario is exemplified in Fig. 9. Assuming a trigonal distortion of the octahedral crystal field, the excited quartet will be further split into two doublets, resulting in three Kramers doublets as shown in Fig. 9(c). So far, from Raman [44] and neutron scattering experiments [25] it has been assumed that the spin-orbit splitting Δ amounts 150 – 200 meV. No indications of a second transition were reported. Values of the trigonal splitting, here $\Delta_2 - \Delta_1$, range between 10 and 100 meV [10,71,72]. The fact that only one transition is observable was interpreted as being due to an almost cubic symmetry with a very small trigonal distortion [72]. In the related Kitaev-type Iridates with stronger spin-orbit coupling, onsite d-d excitations have been observed well below 1 eV and the trigonal splitting was estimated as being ~ 100 meV [73].

With the level scheme indicated in figure 9 and with the excitation energies indicated above, one interpretation would be that $\Delta_1$ ~ 70 meV and $\Delta_2$ ~ 250 meV correspond to excitations between the three Kramers doublets. Clearly these values are not compatible with results reported so far and would indicate a relatively strong trigonal splitting not in accord with recent x-ray absorption spectroscopy results [72]. In addition, the origin of the remaining three excitation energies remains unclear. So far, the remaining excitations have to be interpreted in terms of $t_{2g}^5 \rightarrow t_{2g}^4 e_g^1$ excitations, which however, is not supported by theory: from quantum chemistry electronic structure calculations, these excitations are expected well beyond 1 eV, which is true in *C2/m* as well as in *P3$_1$12* symmetry [71]. In the latter symmetry, Yadav et al. [71] even calculated the $t_{2g}^5 \rightarrow t_{2g}^3 e_g^2$ transition at 920 meV, well below excitations from the ground state to the $t_{2g}^4 e_g^1$ levels. It also is far from being settled, if one of the remaining peaks corresponds to the charge gap. While the majority of researchers identify the optical gap of α-RuCl$_3$ above 1 eV, dc conductivity experiments point towards a gap of ~ 100 meV and the onset of the first absorption peak at ~ 200 meV has been identified as possible charge gap [13,51]. One has to keep in mind that the level scheme as outlined in figure 9 corresponds to a purely local picture of on-site excitations. This description strictly holds only in the case of Ru$^{3+}$ diluted in an isolating host. α-RuCl$_3$ is a strongly correlated material and it is unclear how a band picture would change these atomic-like excitations.

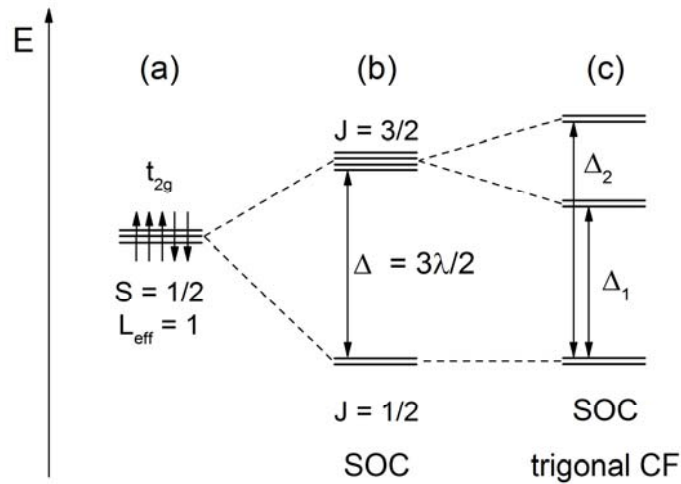

Figure 9. Spin-orbital excitation spectrum of α-RuCl$_3$ in an octahedral crystal field. (a): Low spin t$_{2g}$ ground state, forming a state with an effective angular momentum of L$_{eff}$ = 1 and spin S = ½. (b) By spin-orbit coupling this ground state splits into a lower doublet and an excited quartet. (c) By a further trigonal distortion this SOC state splits into three Kramers doublets. In recent literature estimates of Δ are of order 200 meV, while values of the trigonal splitting Δ$_2$ − Δ$_1$ range from ~ 10 to 100 meV (see text).

In addition, we detected an electronic transition close to 400 meV. This excitation only appears on heating in a narrow temperature range, exactly at the transition from the rhombohedral into the monoclinic structure. This excitation exists in neither the low- nor the high-temperature phase. This excitation could corresponds to a spin and parity forbidden transition, which becomes disorder-allowed just in the temperature range of the structural transformation. A more appealing interpretation would be the hypothesis that these extra excitations correspond to drastic changes in the electronic density of states when during the structural phase transition neighboring molecular stacks are shifted, similar to recent observations in graphene when twisting neighboring layers [17,18]

We conclude and have to admit that these detailed optical experiments on α-RuCl$_3$ created more open questions than have been solved. From our point of view, the striking excitations in the THz regime, which could be remnants of Kitaev physics and the completely unclear sub-gap excitation spectrum < 1 eV, deserve further theoretical attention. Not to forget the appearance of an electronic transition in a very narrow temperature range just across the structural phase transition. Finally, this type of structural phase transition, characterized by a mere switching of the stacking sequence of molecular layers, which we studied here in α-RuCl$_3$, is a characteristic feature of a large number of layered tri-halides. It certainly would be interesting to study in detail this exotic phase transition in isostructural compounds.

**Acknowledgements**

We thank Philipp Gegenwart, Roser Valenti, Stephen Winter, and Igor Mazin for stimulating discussions and helpful comments. This research was partly funded by the Deutsche Forschungsgemeinschaft DFG via the Transregional Collaborative Research Center TRR 80 "From Electronic correlations to functionality" (Augsburg, Munich, Stuttgart). KYC was supported by Korea Research Foundation (KRF) Grants (No.: 2017012642). The work of MVE was funded by a grant, allocated to the Kazan Federal University for the state assignment in the sphere of scientific activities (Contract number: #3.6722.2017/8.9).

**ORCID iDs**

Alois Loidl  https://orcid.org/0000-0002-5579-0746